\documentclass[aps,prb,twocolumn,showpacs,preprintnumbers,amsmath,amssymb,longbibliography]{revtex4-2}

\newcommand{\expect}[1]{\langle #1 \rangle}

\newcommand{\be}{\begin{equation}}
\newcommand{\ee}{\end{equation}}
\newcommand{\bea}{\begin{eqnarray}}
\newcommand{\eea}{\end{eqnarray}}
\newcommand{\eq}[1]{Eq.~(\ref{#1})}
\newcommand{\fig}[1]{Fig.~\ref{#1}}

\newcommand{\e}{\varepsilon}
\newcommand{\w}{\omega}
\newcommand{\s}{\sigma}
\newcommand{\G}{\Gamma}
\newcommand{\up}{\uparrow}
\newcommand{\down}{\downarrow}

\newcommand{\exch}{\Delta\e_{\rm exch}}
\newcommand{\Sn}{S_{\rm n}}
\newcommand{\Sd}{S_{\rm d}}

  \newcommand{\Sec}[1]{Sec.~\ref{#1}}
  
  \newcommand{\Eq}[1]{Eq.~\eqref{#1}}

  \newcommand{\Fig}[1]{Fig.~\ref{#1}}

\usepackage{float}
\usepackage[dvipsnames]{xcolor}
\usepackage{ulem}
\usepackage{verbatim} 
\usepackage{amsmath}
\usepackage{graphicx}% Include figure files
\usepackage{dcolumn}% Align table columns on decimal point
\usepackage{bm}% bold math
\usepackage{hyperref}% add hypertext capabilities

\definecolor{oucrimsonred}{rgb}{0.6, 0.0, 0.0}

\begin{document}

%\title{Thermoelectrics of an asymmetric quantum dot far-from-equilibrium}
%\title{Spin effects on nonequilibrium Seebeck effect in nanoscale junctions}
%\title{Nonequilibrium thermopower and thermoelectric efficiency of an asymmetrically coupled quantum dot spinvalve}
%\title{Spin effects on nonequilibrium thermopower of correlated nanoscale junctions}
\title{Nonequilibrium Seebeck and spin Seebeck effects in nanoscale junctions}

\author{Anand Manaparambil}
\email{anaman@amu.edu.pl}
\affiliation{Institute of Spintronics and Quantum Information,
Faculty of Physics, 
Adam Mickiewicz University,
Uniwersytetu Pozna\'nskiego 2, 61-614 Pozna\'n, Poland}

\author{Ireneusz Weymann}
\affiliation{Institute of Spintronics and Quantum Information,
Faculty of Physics, 
Adam Mickiewicz University,
Uniwersytetu Pozna\'nskiego 2, 61-614 Pozna\'n, Poland}

\date{\today}

%%%%%%%%%%%%%%%%%%%%%%%%%%%%%%%%%%%%%%%%%%%%%%%%%%
%%%%%%%%%%%%%%%%%%%%%%%%%%%%%%%%%%%%%%%%%%%%%%%%%%

\begin{abstract}
The spin-resolved thermoelectric transport properties of correlated nanoscale junctions,
consisting of a quantum dot/molecule asymmetrically coupled to external ferromagnetic contacts,
are studied theoretically in the far-from-equilibrium regime.
One of the leads is assumed to be strongly coupled to the quantum dot
resulting in the development of the Kondo effect.
The spin-dependent current flowing through the system, as well as the thermoelectric properties,
are calculated by performing a perturbation expansion with respect to the
weakly coupled electrode, while the Kondo correlations are captured accurately
by using the numerical renormalization group method.
In particular, we determine the differential and nonequilibrium Seebeck effects
of the considered system in different magnetic configurations
and uncover the crucial role of spin-dependent tunneling on the device performance.
Moreover, by allowing for spin accumulation in the leads,
which gives rise to finite spin bias,
we shed light on the behavior of the nonequilibrium spin Seebeck effect.
\end{abstract}

\maketitle

%%%%%%%%%%%%%%%%%%%%%%%%%%%%%%%%%%%%%%%%%%%%%%%%%%
\section{\label{sec:level1}Introduction}
%%%%%%%%%%%%%%%%%%%%%%%%%%%%%%%%%%%%%%%%%%%%%%%%%

Quantum transport through nanoscale systems, such as quantum dots,
molecular junctions and nanowires, has been under tremendous
research interest due to promising applications of such nanostructures in nanoelectronics,
spintronics and spin-caloritronics \cite{Zutic2004Apr, Bauer2012May, Awschalom2013Mar, Hirohata2020Sep}.
Due to the strong electron-electron interactions and a characteristic discrete density of states, these systems
can exhibit large thermoelectric figure-of-merit and are excellent candidates
for nanoscale heat engines \cite{Dhar2008Sep, Dubi2009Jan, Dubi2011Mar, Benenti2017Jun,Josefsson2018Oct}.
As far as more fundamental aspects are concerned,
correlated nanoscale systems allow one to explore fascinating
many-body phenomena that are not present in bulk materials.
One of such phenomena is the Kondo effect, which can drastically change
the system's transport properties at low temperatures by giving rise to 
a universal enhancement of the conductance to its maximum \cite{Kondo1964Jul, Hewson_1993, Costi2010Jun}.
Moreover, in addition to voltage-biased setups' investigations, the emergence of Kondo correlations can be probed
in the presence of a temperature gradient, where thermoelectric transport properties reveal the important physics \cite{Dhar2008Sep, Dubi2009Jan, Dubi2011Mar}.
In fact, the thermopower of the quantum dot systems
have been shown to contain the signatures of the Kondo effect.
Specifically, the sign changes in the temperature dependence of the thermopower
with the onset of Kondo correlations have been identified in both the theoretical \cite{Costi2010Jun}
and experimental \cite{Svilans2018Nov, Dutta2019Jan,Hsu2022Apr} studies.

Furthermore, other interesting properties arise when the electrodes are magnetic,
making such nanoscale systems important for spin nanoelectronics applications \cite{Awschalom2013Mar,Hirohata2020Sep}.
It turns out that ferromagnetism of the leads can compete with the Kondo correlations
giving rise to an interplay between ferromagnet-induced exchange
field and the Kondo behavior \cite{Martinek2003Dec, Pasupathy2004Oct, Hamaya2007Dec,Weymann2011Mar}.
This interplay has been revealed in theoretical studies
on thermoelectric properties of strongly-correlated molecular and 
quantum dot systems with ferromagnetic contacts \cite{Krawiec2006Feb,Weymann2013Aug}.
From theoretical point of view, accurate description of
low-temperature transport behavior of correlated nanoscale systems
with competing energy scales requires resorting to advanced numerical methods,
such as the numerical renormalization group (NRG) method
\cite{Wilson1975Oct,Bulla2008Apr}.
Indeed, while there has been a tremendous progress in complete understanding
of transport properties at equilibrium \cite{,Karwacki2013Nov,
Wojcik2016Feb,Karwacki2016Aug,Wojcik2016Dec,
Gorski2018Sep,Manaparambil2021Apr,Majek2022Feb},
much less is known in fully nonequilibrium settings, where standard NRG cannot be applied. 
The exact treatment of the nonlinear response regime requires even
more sophisticated numerical techniques \cite{Schwarz2018Sep,Manaparambil2022Sep}
and this is why it has been much less explored
\cite{Sierra2014Sep,Svilans2016Dec,Sierra2017Aug,Khedri2018Nov,Eckern2020Jan,Manaparambil2023Feb}.

In this work we therefore investigate the nonlinear thermopower
of a molecular magnetic junction and analyze how the spin-resolved
transport affects the nonequilibrium thermoelectric properties of the system.
More specifically, we consider a quantum dot/molecule
strongly coupled to one ferromagnetic lead
and weakly coupled to the other nonmagnetic or ferromagnetic
lead kept at different potentials and temperatures, see Fig.~\ref{fig:schem}.
We perform a perturbation expansion in the weak coupling,
while the strongly coupled subsystem, where Kondo correlations
may arise, is solved with the aid of the NRG method.
This allows us to extract the signatures of the interplay between the spin-resolved transport
and the Kondo correlations in the Seebeck coefficient in far from equilibrium settings.
Furthermore, we study how different magnetic configurations of the system
affect the differential and nonequilibrium Seebeck effects.
In particular, we show that the Seebeck effect exhibits new sign changes 
as a function of the bias voltage which are associated with
the Kondo resonance split by exchange field.
These sign changes are found to extend to the temperature
gradients on the order of the Kondo temperature.
Moreover, we also provide a detailed analysis of the
nonequilibrium spin Seebeck coefficient. 
We believe that our work sheds light on the spin-resolved
nonequilibrium thermopower of correlated nanoscale junctions,
in which the interplay between the Kondo and exchange field correlations is relevant. 
It thus provides a better understanding of spin caloritronic nanodevices
under finite temperature and voltage gradients.

The paper is organized as follows:
The system Hamiltonian and the theoretical framework are described in \Sec{sec:theory}.
The numerical calculations and the results are discussed in \Sec{sec:result}
with a short summary and concluding remarks in \Sec{sec:summary}.

%%%%%%%%%%%%%%%%%%%%%%%%%%%%%%%%%%%%%%%%%%%%%%%%%%
\section{\label{sec:theory}Theoretical description}
%%%%%%%%%%%%%%%%%%%%%%%%%%%%%%%%%%%%%%%%%%%%%%%%%%
%%%%%%%%%%%%%%%%%%%%%%%%%%%%%%%%%%%%%%%%%%%
\subsection{Hamiltonian of the system}
%%%%%%%%%%%%%%%%%%%%%%%%%%%%%%%%%%%%%%%%%%%

\begin{figure}[t]
 	\includegraphics[width=0.9\columnwidth]{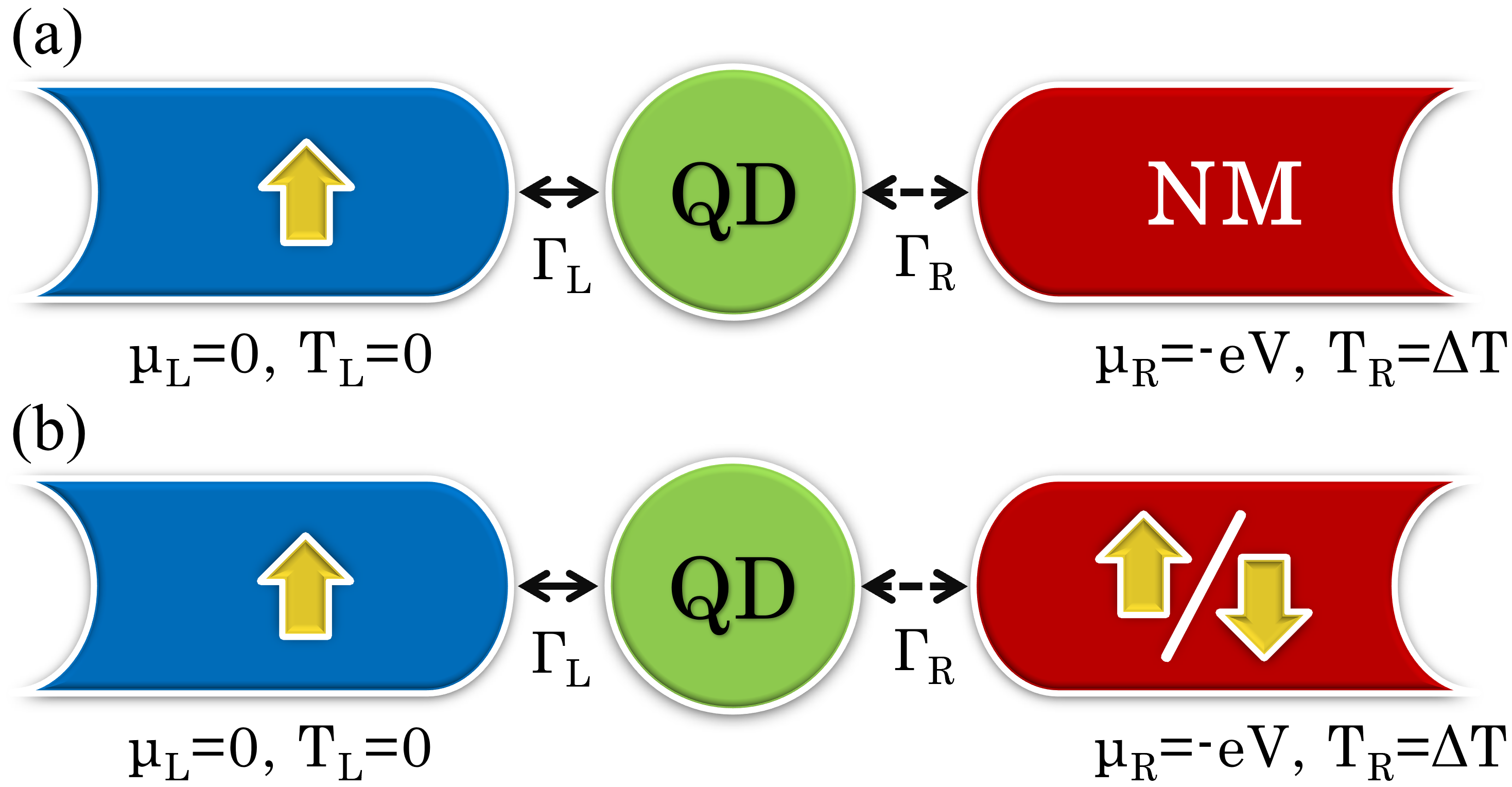}
 	\caption{The schematic of the considered asymmetric tunnel junction
 		with embedded quantum dot/molecule strongly coupled to
 		a cold ferromagnetic left lead and weakly coupled to
 		a hot (a) nonmagnetic or (b) ferromagnetic right lead.
 		The right lead is subject to voltage and temperature gradients,
 		while the left lead is grounded and kept at zero temperature.
 		The device in (b) can be in two magnetic configurations:
 		the parallel (P) and antiparallel (AP) one, as indicated by the arrows.}
    \label{fig:schem}
\end{figure}

We consider a nanoscale junction with an embedded
quantum dot/molecule, which is schematically shown in \Fig{fig:schem}.
The quantum dot is assumed to be strongly coupled to the left ferromagnetic lead
and weakly coupled to the right lead, which can be either nonmagnetic [\Fig{fig:schem}(a)]
or ferromagnetic [\Fig{fig:schem}(b)]. In the case of two ferromagnetic electrodes,
we will distinguish two magnetic configurations: the parallel (P) one
when the leads magnetic moments point in the same direction
and the antiparallel (AP) one,
when the orientation of magnetic moments is opposite, see \Fig{fig:schem}(b).
It is assumed that there are finite temperature and voltage gradients applied to the system,
with $T_L=0$ and $\mu_L=0$, whereas $T_R=\Delta T$ and $\mu_R=-eV$,
as shown in \Fig{fig:schem}, where $T_\alpha$ and $\mu_\alpha$ are 
the temperature ($k_B\equiv 1$) and the chemical potential of lead $\alpha$.

With the assumption of weak coupling between the quantum dot and right contact
the system Hamiltonian can be simply written as
\begin{equation}
	H=H_L + H_R + H_T.
\end{equation}
$H_L$ describes the strongly coupled left subsystem,
consisting of the quantum dot and the left lead, and it is given by
\bea
H_L= \e_d\sum_\s n_{\s} + U n_\up n_\down + \sum_{k\s} \e_{Lk\s} c^\dagger_{Lk\s} c_{Lk\s} \nonumber \\
+ \sum_{k\s} %
t_{Lk\s}
 (d^\dagger_\s c_{Lk\s} + c^\dagger_{Lk\s} d_\s),
\eea
where $n_\s=d^\dagger_\s d_\s$, with $d^\dagger_\s$ ($d_\s$) being
the creation (annihilation) operator on the quantum dot for an electron of spin $\s$,
$c_{\alpha k\sigma}$ ($c_{\alpha k\sigma}^\dag$) annihilates (creates)
an electron in the lead $\alpha$ with momentum $k$, spin $\sigma$
and energy $\e_{\alpha k \s}$. The quantum dot is modeled by a single orbital of energy
$\e_d$ and Coulomb correlations $U$. The hopping matrix elements between the quantum dot
and lead $\alpha$ are denoted by $t_{\alpha k \s}$ and give rise to the level broadening
$\Gamma_{\alpha\s} = \pi\rho_{\alpha \s} |t_{\alpha k \s}|^2$, which is
assumed to be momentum independent, where $\rho_{\alpha \s}$
is the density of states of lead $\alpha$ for spin $\sigma$.

The second part of the Hamiltonian describes the right lead and is given by
\begin{equation}
	H_R = \sum_{k\s} \e_{Rk\s} c^\dagger_{Rk\s} c_{Rk\s} - e \sum_{k\s} \mu_{R\s} c^\dagger_{Rk\s} c_{Rk\s},
\end{equation}
while the last term of $H$ accounts for the hopping between the left and right subsystems
\begin{equation}
	H_T = \sum_{k\s} t_{Rk\s} (d^\dagger_\s c_{Rk\s} + c^\dagger_{Rk\s} d_\s).
\end{equation}
In the following, we use the lowest-order perturbation theory in $H_T$
to study the spin-dependent electric and thermoelectric properties of the system.

%%%%%%%%%%%%%%%%%%%%%%%%%%%%%%%%%%%%%%%%%%%%%%%%%%%%%%
\subsection{Nonlinear transport coefficients}
%%%%%%%%%%%%%%%%%%%%%%%%%%%%%%%%%%%%%%%%%%%%%%%%%%%%%%

The electric current flowing through the system in the spin channel $\s$ can be expressed as
\cite{Csonka2012May,Tulewicz2021Jul}
\begin{eqnarray}
I_{\s} (V,\Delta T) &=& -\frac{e\G_{R \s}}{\hbar}\int_{-\infty}^{\infty}d\omega \; A_{L\s} (\omega) \nonumber \\
&&\times  [f_L (\omega) - f_R (\omega-eV)],
\label{Eq:I}
\end{eqnarray}
where $f_{\alpha} (\omega) = [1+{\rm exp} (\omega/T_{\alpha})]^{-1}$ 
is the Fermi-Dirac distribution function, while 
$A_{L\s}(\omega)$ denotes the spin-resolved spectral function of the left subsystem.
The total current flowing through the system under potential bias $V$
and temperature gradient $\Delta T$ is thus $I (V,\Delta T) =\sum_\s I_\s (V,\Delta T) $.
The spectral function $A_{L\s}(\omega)$ is calculated by means of the NRG method
\cite{Wilson1975Oct,Bulla2008Apr,FlexibleDMNRG},
which allows us to include all the correlation effects between the quantum dot strongly coupled
to left contact in a fully nonperturbative manner. In particular,
$A_{L\s}(\omega)$ is determined as the imaginary part of the Fourier transform
of the retarded Green's function of the left subsystem Hamiltonian $H_L$,
${G_{\s}(t) = -i\Theta(t)\expect{\{ d_\s (t), d_\s^\dag(0) \}}}$.
In NRG calculations, the spectral data is collected in logarithmic bins
that are then broadened to obtain a smooth function.

For the further analysis, it is convenient to express
the coupling constants $\Gamma_{\alpha \s}$ by using the spin polarization of the lead $\alpha$, $p_\alpha$,
as $\Gamma_{L \s} = (1+\s p_L) \Gamma_L$ and $\Gamma_{R \s} = (1+\s p_R)\Gamma_R$
for the parallel magnetic configuration, with $\Gamma_{R \s} = (1- \s p_R) \Gamma_R$
in the case of the antiparallel configuration of the system.
Here, $\Gamma_\alpha = (\Gamma_{\alpha \uparrow} + \Gamma_{\alpha \downarrow})/2$.
Furthermore, in the case when the right lead is nonmagnetic, $p_R = 0$,
while for both ferromagnetic leads we for simplicity assume $p_L=p_R\equiv p$.

As far as thermoelectric coefficients are concerned,
the differential Seebeck coefficient can be expressed as \cite{Dorda2016Dec}
\be
S_d=-\left( \frac{dV}{d \Delta T}\right)_{\!\! I}
=-\left(\frac{\partial I}{\partial \Delta T}\right)_{\!\!V}\bigg/\!\! \left( \frac{\partial I}{\partial \Delta T}\right)_{\!\! \Delta T}.
\ee
Furthermore, the extension of the conventional Seebeck coefficient to the nonlinear response regime
is referred to as the nonequilibrium Seebeck coefficient $S_n$, and it can be defined as
\cite{Krawiec2007Apr,Leijnse2010Jul,Azema2014Nov,Erdman2017Jun,Eckern2020Jan,PerezDaroca2018Apr}
\be
S_n=-\left(\frac{\Delta V}{\Delta T}\right)_{\!\! I(V+\Delta V,\Delta T)=I(V,0)}.
\ee

The above definitions will be used to describe thermoelectric transport
in different configurations of the system, respectively.

%%%%%%%%%%%%%%%%%%%%%%%%%%%%%%%%%%%%%%%%%%%%%%%%%%
\section{\label{sec:result}Numerical results and discussion}
%%%%%%%%%%%%%%%%%%%%%%%%%%%%%%%%%%%%%%%%%%%%%%%%%%

In this section we present the main numerical results and their discussion.
In our considerations we assume that the left lead is always ferromagnetic,
while the right electrode can be either nonmagnetic or ferromagnetic, cf. Fig.~\ref{fig:schem}.
For the studied setup, the strong coupling to the left contact may give rise to the Kondo effect \cite{Hewson_1993,Goldhaber-Gordon1998Jan}.
However, it is crucial to realize that the presence of the spin-dependent hybridization results in
a local exchange field on the quantum dot, which can split the dot orbital level
when detuned from the particle-hole symmetry point, and thus suppress the Kondo resonance.
The magnitude of such exchange field can be estimated from the perturbation theory,
which at zero temperature gives \cite{Martinek2003Sep},
\be
\exch = \frac{2 p_L \G_L}{\pi} \ln \left| \frac{\e_d}{\e_d+U} \right|.
\label{eq:Exch}
\ee
The presence of the exchange field and its detrimental effect
on the Kondo phenomenon has been confirmed by various experiments
on electronic transport measurements in quantum dot and molecular systems
\cite{Pasupathy2004Oct,Hamaya2007Dec,Hauptmann2008May,Gaass2011Oct}.

We start our considerations with the analysis of electric transport properties,
revealing the effects of the exchange field.
Further on, we study the nonlinear thermoelectric response,
first for the case of nonmagnetic right lead
and then for the case of two ferromagnetic leads.
In numerical calculations, we use the following parameters: $U=0.2$,
$\Gamma_L=0.02$, $\Gamma_R=0.002$, in units of band halfwidth,
and $p=0.4$ for the ferromagnetic leads.
For the assumed parameters, the Kondo temperature
of the left subsystem for $\e_d = -U/2$ is equal to \cite{Haldane1978,Martinek2003Sep},
$T_K \approx 0.035 \Gamma_L$.

\begin{figure}[t]
	\includegraphics[width=0.9\columnwidth]{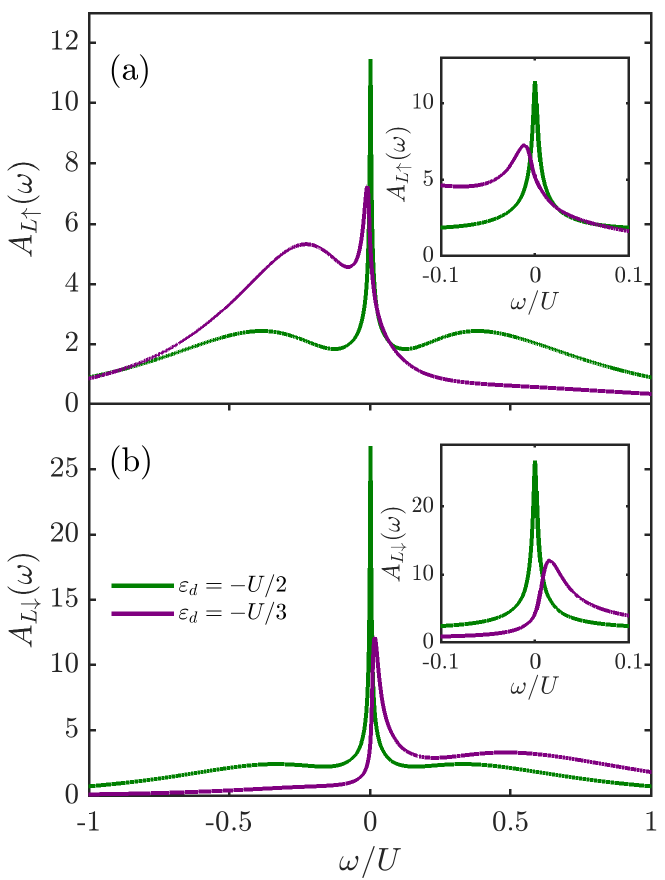}
	\caption{The energy dependence of the spectral functions for the individual spin channels,
		(a) $A_{L\uparrow} (\omega)$ and (b) $A_{L\downarrow} (\omega)$
		calculated for the strongly coupled left subsystem with orbital energies as indicated.
		The zoomed Kondo and split-Kondo peaks are shown in the insets.
		The other parameters are: $U=0.2$, $\Gamma_L=0.02$, in units of band halfwidth, and $p=0.4$.
	}
	\label{FigA}
\end{figure}

To begin with, it is instructive to analyze the properties of the left subsystem itself
as described by the spectral function. The spectral function
for each individual spin channel is shown in \fig{FigA}.
First of all, one can see that for $\e_d=-U/2$ there is a pronounced
Kondo peak at the Fermi level for each spin component.
However, when detuned from the particle-hole symmetry,
there is a finite exchange-induced splitting, cf. Eq.~(\ref{eq:Exch}),
which suppresses the Kondo effect when $|\exch|\gtrsim T_K$,
with $T_K$ denoting the Kondo temperature.
Because of that, each spin component of the spectral function
displays a slightly detuned from Fermi energy side peak,
constituting the split Kondo resonance.
In addition, the Hubbard resonances at $\omega \approx \e_d$ and $\omega \approx \e_d+U$
become affected as well: although their position is only slightly modified,
their magnitude gets strongly spin-dependent.

\begin{figure}[t]
	\includegraphics[width=0.95\columnwidth]{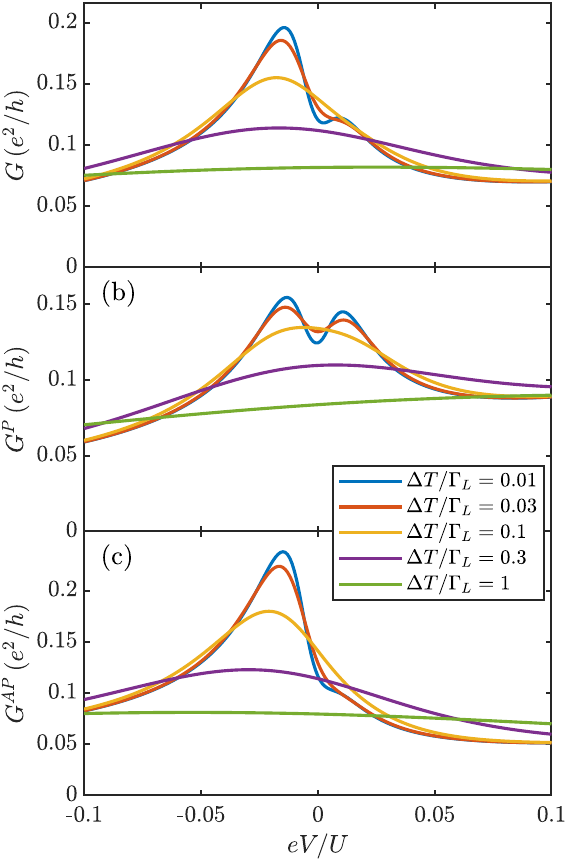}
	\caption{The differential conductance
		for the quantum dot strongly coupled to ferromagnetic
		left lead and weakly coupled to
		(a) nonmagnetic right lead,
		ferromagnetic right lead in (b) the parallel magnetic configuration
		and (c) the antiparallel magnetic configuration.
		The parameters are the same as in \fig{FigA} with $\e_d=-U/3$
		and different temperature gradients, as indicated.
	}
	\label{fig:G}
\end{figure}

The splitting of the Kondo resonance is directly visible
in the differential conductance of the system,
which is demonstrated in Fig.~\ref{fig:G}.
This figure presents the bias voltage dependence of the differential conductance
in different magnetic configurations
for various temperature gradients, as indicated.
More specifically, $G$ corresponds to the case when the right lead is nonmagnetic
[cf. Fig.~\ref{fig:schem}(a)], while $G^P$ ($G^{AP}$) presents the case
of both ferromagnetic leads in the parallel (antiparallel) alignment
[cf. Fig.~\ref{fig:schem}(b)]. 
When the orbital level is detuned out of the particle-hole symmetry point,
$\e_d=-U/3$, as in the case of \fig{fig:G},
the splitting of the Kondo peak in the spectral function
of the left subsystem becomes revealed in the differential conductance
of the whole system. Let us start with the case of nonmagnetic right lead,
presented in \fig{fig:G}(a). First of all, one can note a large asymmetry
of the differential conductance with respect to the bias reversal.
Moreover, for small temperature gradients, 
$\Delta T \lesssim T_K$, the split zero-bias anomaly
due to the Kondo effect is visible.
These features can be understood by inspecting the 
behavior of the spectral function around the Fermi energy,
see the insets in \fig{FigA}. One can note that the split Kondo peak
in $A_{L\uparrow}(\w<0)$ has smaller weight 
compared to the split Kondo peak in $A_{L\downarrow}(\w>0)$.
Because, for low temperature gradients, for $eV>0$ ($eV<0$)
we probe the density of states of the left subsystem for negative (positive) energies,
the above-mentioned asymmetry in $A_{L\sigma}(\w)$
gives rise to highly asymmetric behavior of the differential conductance, see \fig{fig:G}(a),
with the peak in the negative voltage regime more pronounced than the other.
Interestingly, when the tunneling to the right lead
becomes spin dependent, in the case of parallel configuration
one observes a rather symmetric behavior of $G^{P}$,
with nicely visible split zero-bias anomaly, see \fig{fig:G}(b).
This is due to the fact that the increased tunneling rate of spin-down electrons
due to larger density of states becomes now reduced since the spin-down
electrons are the minority ones in the right lead. On the other hand,
the tunneling of spin-up electrons to the right is enlarged.
As a consequence, the unequal contributions of the currents 
in each spin channel become now equalized
and the differential conductance in the parallel configuration
exhibits split-Kondo resonance with the side peaks of comparable height.
On the other hand, when the magnetization of the right
lead is flipped, the asymmetric behavior visible in \fig{fig:G}(a)
is even further magnified, see \fig{fig:G}(c).
This can be understood by invoking similar arguments
as above, keeping in mind that now the rate of spin-up tunneling 
to the right is smaller than that for spin-down electrons.
With increase in the temperature gradient,
the Kondo-related behavior gets smeared
and finally disappears when $\Delta T\gtrsim T_{\rm K},|\exch|$.

%%%%%%%%%%%%%%%%%%%%%%%%%%%%%%%%%%%%%%%%%%%%%%%%%%%%%%
\subsection{Effects of exchange field on nonequilibrium thermopower}
%%%%%%%%%%%%%%%%%%%%%%%%%%%%%%%%%%%%%%%%%%%%%%%%%%%%%%

In this section, we focus on the case where the right lead is nonmagnetic,
see Fig.~\ref{fig:schem}(a). In such a setup it will be possible to observe
clear signatures of ferromagnet-induced exchange field
on the thermoelectric properties of the system subject to 
temperature and voltage gradients.
We first study the case of the linear response in potential bias
with nonlinear temperature gradient in Sec.~\ref{sec:nm0},
while in Sec.~\ref{sec:nm_nl} the discussion is extended
to the case of nonlinear response regime in both $\Delta T$ and $V$.

%%%%%%%%%%%%%%%%%%%%%%%%%%%%%%%%%%%%%%%%%%%%%%%%%%%%%%
\subsubsection{Zero-bias thermoelectrics with finite temperature gradient \label{sec:nm0}}
%%%%%%%%%%%%%%%%%%%%%%%%%%%%%%%%%%%%%%%%%%%%%%%%%%%%%%

\begin{figure}[t]
 	\includegraphics[width=0.95\columnwidth]{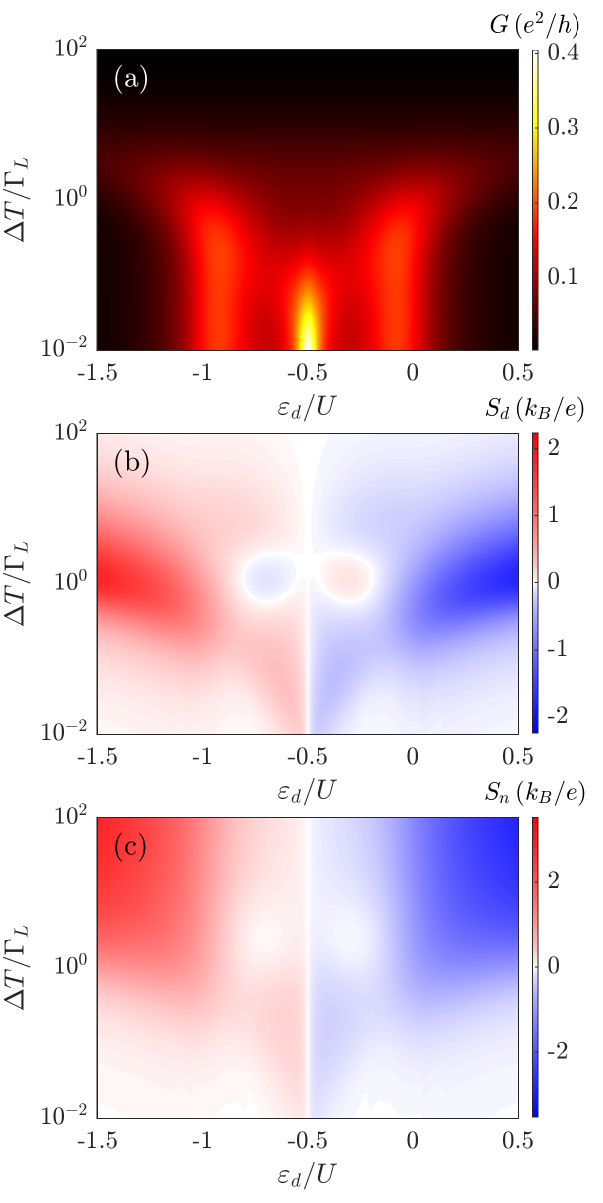}
 	\caption{(a) The differential conductance $G$,
 		(b) the differential Seebeck coefficient $S_d$
 		and (c) the nonequilibrium Seebeck coefficient $S_n$
 		of the quantum dot strongly coupled to left ferromagnetic lead
 		and weakly attached to the right nonmagnetic lead plotted as a function of the orbital energy $\e_d$
 		and the temperature gradient $\Delta T$. The system is assumed
 		to be in the linear response regime with respect to the bias voltage.
 		The other parameters are the same as in \fig{FigA}.}
    \label{fig:lin_NM}
\end{figure}

Figure \ref{fig:lin_NM} displays the zero-bias differential conductance $G$,
the differential Seebeck coefficient $\Sd$ and the nonlinear
Seebeck coefficient $\Sn$ calculated as a function of orbital level $\e_d$
and finite temperature gradient $\Delta T$.
For low temperature gradients, the conductance
shows considerable increase near three values of $\e_d$.
The peaks for $\e_d \approx 0$ and $\e_d\approx -U$ 
correspond to the Hubbard resonances in the spectral function,
whereas the maximum at $\e_d=-U/2$ is due to the Kondo effect.
In fact, in the local moment regime, $-1\lesssim \e_d/U\lesssim 0$,
the Kondo resonance is suppressed by the exchange field once $|\exch|\gtrsim T_K$,
i.e. for values of $\e_d$ away from the particle-hole symmetry point, cf. \eq{eq:Exch}.
With the increase in the temperature gradient, the Kondo resonance dies out when
$\Delta T>T_K$ and the Hubbard peaks get suppressed when $\Delta T > \G_L$,
see \fig{fig:lin_NM}(a).

In the case of differential and nonlinear Seebeck coefficients
presented in Figs.~\ref{fig:lin_NM}(b) and (c), respectively,
we can see an overall antisymmetric behavior across
the particle-hole symmetry point $\e_d=-U/2$.
The sign of the Seebeck coefficient here corresponds
to the dominant charge carriers in transport,
holes for $\e_d<-U/2$ and particles for $\e_d>-U/2$.
The differential Seebeck coefficient shows two sign
changes in the local moment regime as a function of the temperature gradient.
The sign change around $\Delta T \approx T_K$
originates from the signatures of the Kondo correlations present in the spectral functions.
In the case of nonlinear Seebeck coefficient,
we do not find the corresponding sign changes because
$\Sn$ can deviate considerably from the linear
response Seebeck coefficient at large $\Delta T$ \cite{Manaparambil2021Apr}.
Additionally, one can see that both Seebeck coefficients
decay with decreasing $\Delta T$, this behavior can be described
using the Sommerfeld expansion for the linear response Seebeck coefficient, where 
\be
\left. S(T) \propto \frac{T}{A(\omega=0,T)}\frac{\partial A}{\partial \omega}\right\vert_{\omega=0}.
\ee
We also note that both Seebeck coefficients can possess finite values at even
lower $\Delta T$ inside the local moment regime than
outside of it due to the additional contribution of the Kondo resonance
in the spectral function $A_L(\omega)$ at $\omega=0$.

%%%%%%%%%%%%%%%%%%%%%%%%%%%%%%%%%%%%%%%%%%%%%%%%%%%%%%
\subsubsection{The case of nonlinear potential bias and temperature gradients \label{sec:nm_nl}}
%%%%%%%%%%%%%%%%%%%%%%%%%%%%%%%%%%%%%%%%%%%%%%%%%%%%%%

\begin{figure*}[t]
 	\includegraphics[width=0.9\textwidth]{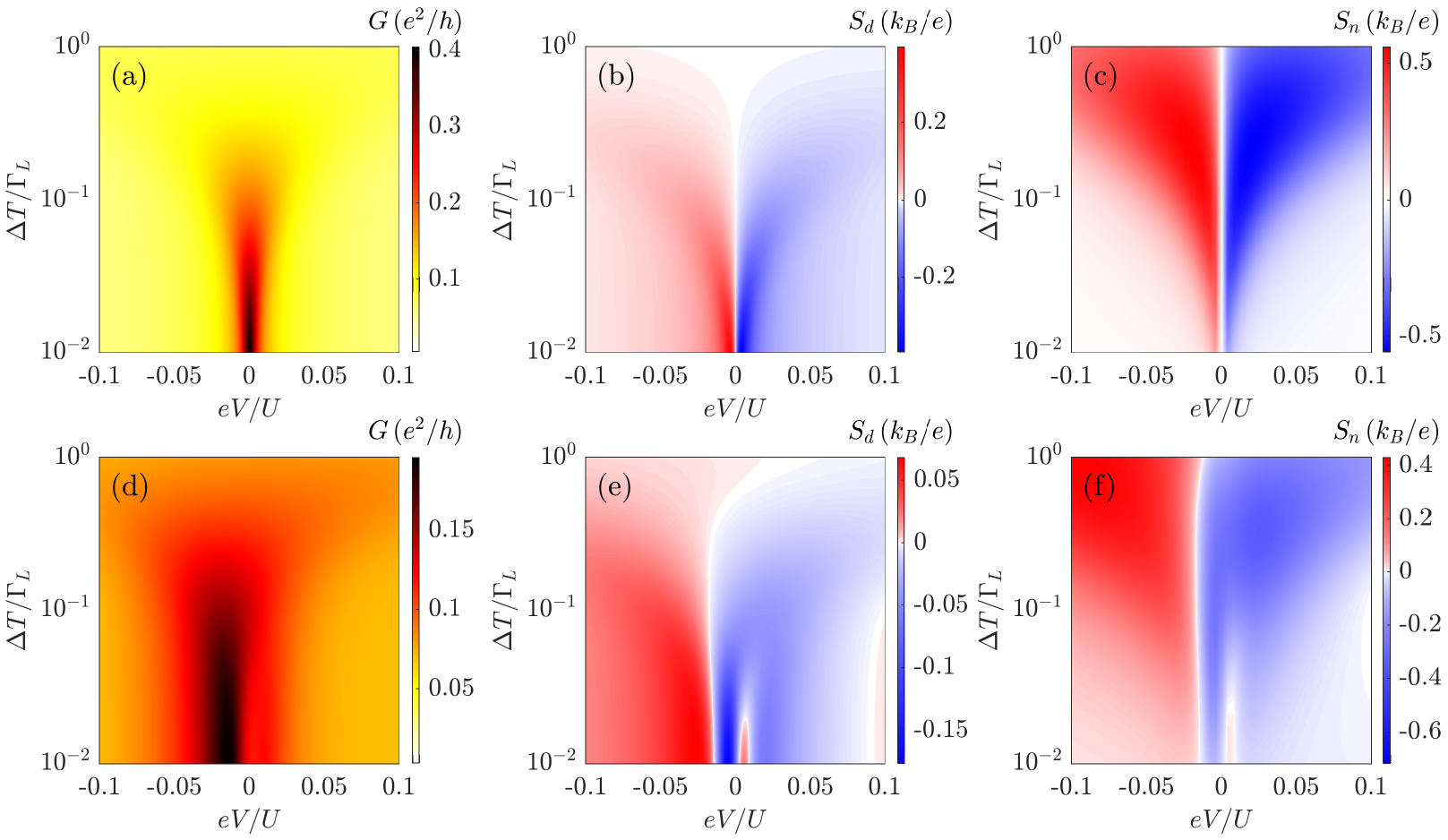}
 	\caption{(a,d) The differential conductance $G$, (b,e) the differential Seebeck coefficient $S_d$
 		and (c,f) the nonequilibrium Seebeck coefficient $S_n$ as 
 		a function of the potential bias $V$ and the temperature gradient $\Delta T$.
 		The first row corresponds to the particle-hole symmetry
 		point $\e_d=-U/2$, while the second row shows the case of $\e_d=-U/3$.
 		The other parameters are the same as in \fig{fig:lin_NM}.
 		}
    \label{fig:nonlin_NM}
\end{figure*}

Let us now inspect the behavior of the nonequilibrium thermoelectric coefficients
as a function of both potential bias and temperature gradient
shown in \fig{fig:nonlin_NM}, focusing on the $V$ and $\Delta T$ range
where Kondo correlations are important. The first row of the figure corresponds to the
case of particle-hole symmetry, $\e_d = -U/2$, while the second row
presents the results for $\e_d = -U/3$.
Consider the first case. Figure~\ref{fig:nonlin_NM}(a) depicts
the bias and temperature gradient dependence of the differential conductance $G$.
There exist a prominent peak at low $\Delta T$ centered at $V=0$,
this is the zero-bias conductance peak characteristic of the Kondo effect.
As the temperature gradient increases, the Kondo peak dies out
and becomes smeared when $\Delta T\gtrsim T_K$.
The differential and nonlinear Seebeck coefficients,
shown in Figs.~\ref{fig:nonlin_NM}(b) and (c),
exhibit a sign change with respect to the bias voltage reversal.
Moreover, while $S_d$ exhibits considerable values
around the Kondo peak and becomes suppressed as $\Delta T$ grows,
$S_n$ gets enhanced when $\Delta T \gtrsim (\G_L/U)|eV|$.

When the orbital level is detuned out of the particle-hole symmetry point,
one can observe an interesting interplay between the exchange field and Kondo effect,
and its signatures present in the nonlinear thermoelectric coefficients.
First, \fig{fig:nonlin_NM}(d) shows the splitting of the Kondo peak
due to the exchange field present in the strongly correlated subsystem.
As observed in the discussions of \fig{fig:G}(a),
the split Kondo peaks are not symmetric,
with the more prominent one in the $eV<0$ regime and both dying off at large $\Delta T$.
Interestingly, the differential and nonlinear Seebeck coefficients
also capture the signatures of the exchange field shown by the split Kondo peak.
In fact, there exist additional sign changes in the
nonlinear response regime with respect to $V$.
More specifically, at low $\Delta T$, there is a sign change at low bias voltages,
followed by another one, roughly located around the split-Kondo peak,
see Figs.~\ref{fig:nonlin_NM}(e) and (f).
These sign changes correspond to the additional energy scale in the system,
namely the exchange field $\exch$. They occur at slightly different
absolute values of $eV$, which is due to the fact that the Kondo resonance
in the local density of states of the left subsystem exhibits
an asymmetric splitting, cf.~\fig{FigA}.
With increasing the temperature gradient, we observe that the right split Kondo peak
in the conductance dies out first, accordingly the regime of positive values
of the Seebeck coefficients corresponding to the right peak disappears around $\Delta T \approx 0.03 \, \G_L$.
Moreover, we also note that the overall sign change of the thermopower
as a function of the bias voltage is now shifted to negative values of $eV$,
as compared to the case of particle-hole symmetry, 
see Fig.~\ref{fig:nonlin_NM}.

\begin{figure*}[t]
	\includegraphics[width=0.8\textwidth]{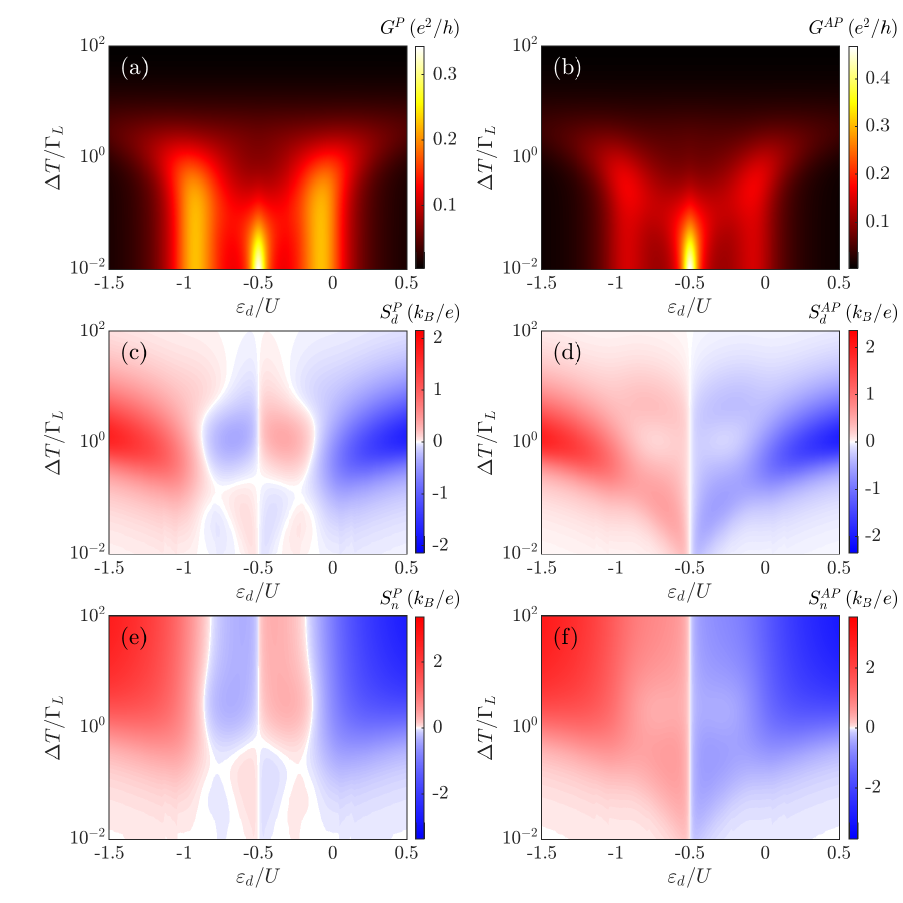}
	\caption{(a,b) The differential conductance $G$,
		(c,d) the differential Seebeck coefficient $S_d$
		and (e,f) the nonequilibrium Seebeck coefficient $S_n$
		in (first column) the parallel (P) and (second column) antiparallel (AP) configuration
		calculated as a function of $\Delta T$ and $\e_d$ assuming linear response in voltage.
		The spin polarizations of both leads are equal to $p=0.4$
		and the other parameters are the same as in \fig{fig:lin_NM}.
	}
	\label{fig:lin_P}
\end{figure*}

%%%%%%%%%%%%%%%%%%%%%%%%%%%%%%%%%%%%%%%%%%%%%%%%%%%%%%
\subsection{Effects of different magnetic configurations on nonequilibrium thermopower}
%%%%%%%%%%%%%%%%%%%%%%%%%%%%%%%%%%%%%%%%%%%%%%%%%%%%%%

In this section we study the case where the quantum dot is coupled to both ferromagnetic leads
with spin polarization $p=0.4$. The magnetic moments
of the external leads are assumed to be aligned either in parallel or antiparallel.
The focus is on the effects of different magnetic configurations
on nonequilibrium thermoelectric transport properties.

%%%%%%%%%%%%%%%%%%%%%%%%%%%%%%%%%%%%%%%%%%%%%%%%%%%%%%
\subsubsection{The case of zero bias with nonlinear temperature gradient}
%%%%%%%%%%%%%%%%%%%%%%%%%%%%%%%%%%%%%%%%%%%%%%%%%%%%%%

\begin{figure*}[t]
	\includegraphics[width=0.9\textwidth]{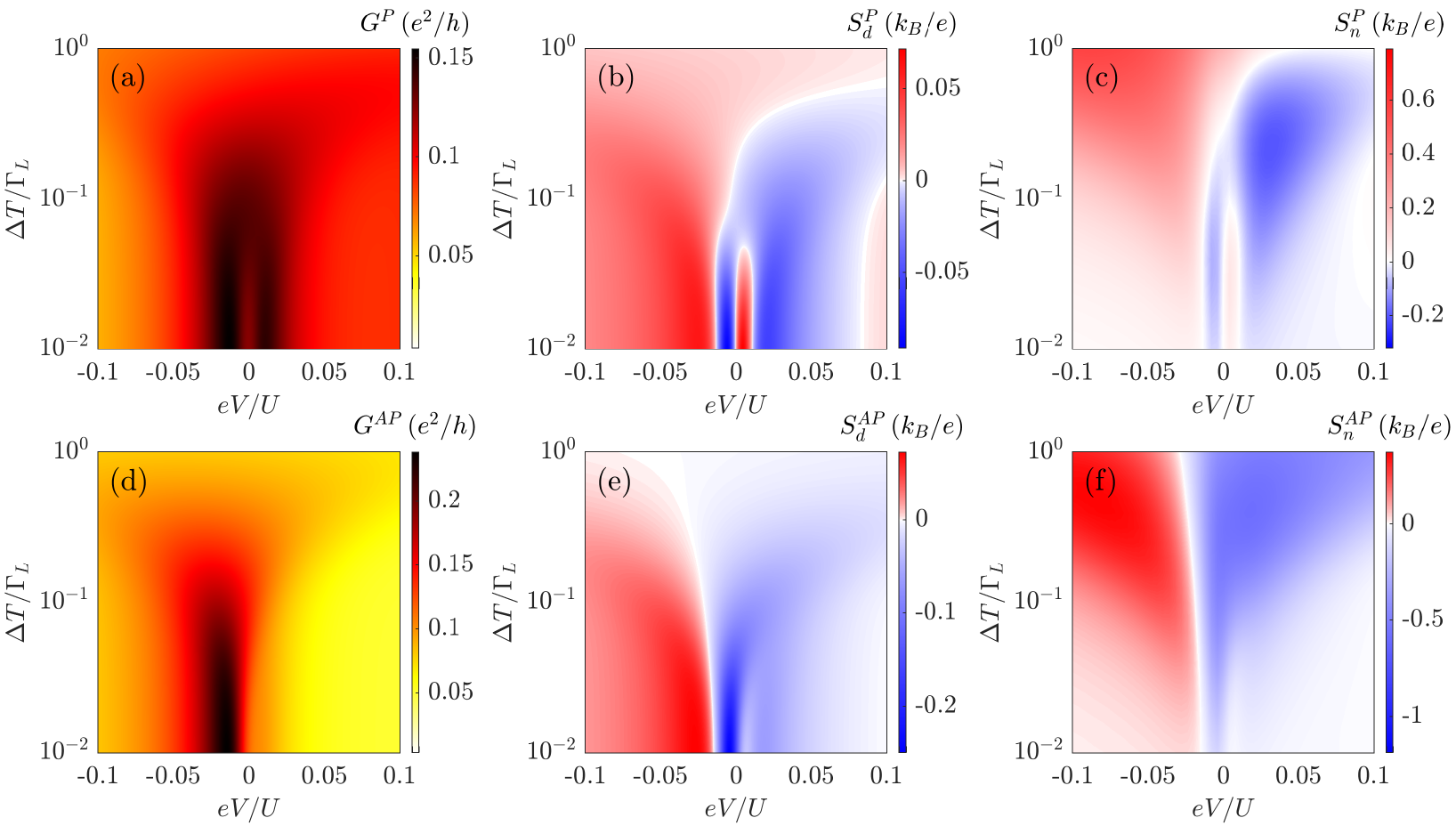}
	\caption{(a,d) The differential conductance $G$,
		(b,e) the differential Seebeck coefficient $S_d$ and
		(c,f) the nonequilibrium Seebeck coefficient $S_n$
		as a function of the bias voltage and temperature greadient
		in the case of $\e_d=-U/3$. The first (second) row corresponds
		to the parallel (antiparallel) magnetic configuration of the system.
		The other parameters are the same as in \fig{fig:lin_P}.
	}
	\label{fig:nonlin_P}
\end{figure*}

The zero-bias thermoelectric properties of the system with two ferromagnetic leads
are shown in \fig{fig:lin_P}. The differential conductance for the parallel $G^{P}$
and antiparallel $G^{AP}$ configuration of the lead magnetizations is shown in 
Figs.~\ref{fig:lin_P}(a) and (b). The qualitative behavior of both conductances
is similar to the case of nonmagnetic lead on the right, where $G$
shows a region of high conductance around $\e_d=-U/2$ due to the Kondo effect.
Similarly to the previous case,
the exchange field suppresses the linear response conductance for values of $\e_d$
away from the particle-symmetry symmetry point.
Around $\e_d \approx 0,-U$, there is a rise in the conductance corresponding to the contribution from the Hubbard peaks.
It is interesting to note that the conductance in the case of parallel configuration
is smaller than that in the antiparallel configuration around the Kondo resonance,
cf. the discussion of Fig.~\ref{fig:G},
while this situation is reversed for the resonances at $\e_d \approx 0,-U$.

The Seebeck coefficients $S_d^P$ and $S_n^P$ shown in \fig{fig:lin_P}(c) and (e)
for the parallel configuration display very interesting features corresponding to various energy scales.
These coefficients show antisymmetric behavior across $\e_d=-U/2$
and sign changes as a function of temperature gradient in the local moment regime $-1\lesssim \e_d/U\lesssim 0$.
Let us first consider the linear response in $\Delta T$ for $S_d^P$.
In this regime one can relate the Seebeck coefficient to the conductance
through the Mott's formula. Thus, the changes of $G^P$ as a function
of orbital level are reflected in the corresponding dependence of the thermopower,
which shows sign changes as $\e_d$ is detuned from the particle-hole symmetry point.
The first sign change occurs when detuning is large enough to induce
the exchange field that suppresses the Kondo effect.
Further sign change occurs at the onset of conductance increase (as function of $\e_d$)
due to the Hubbard resonance. This behavior extends to
higher $\Delta T$ as long as the thermal gradient is smaller than 
the Kondo energy scale (or $\exch$). Otherwise,
another sign change occurs as a function of $\Delta T$, 
see \fig{fig:lin_P}(c). Very similar dependence can be observed
in \fig{fig:lin_P}(e), which shows the nonequilibrium Seebeck coefficient $S_n^P$.
The main difference is present for large $\Delta T$, where $S_n^P$ takes 
considerable values while $S_d^P$ decreases, as explained earlier.

The situation is completely different in the case of the antiparallel configuration,
where one does not see any additional sign changes, neither in $S_d^{AP}$
nor in $S_n^{AP}$, other than the ones present across $\e_d=-U/2$,
see Figs.~\ref{fig:lin_P}(d) and (f).
This can be understood by realizing that the interplay of exchange field
with spin-dependent tunneling to the right contact hinders
the splitting of the Kondo resonance
as a function of the bias voltage.
Consequently, one only observes a single resonance displaced from $V=0$,
cf. \fig{fig:G}(c), which results in much more regular dependence of 
the differential and nonequilibrium Seebeck coefficients.

%%%%%%%%%%%%%%%%%%%%%%%%%%%%%%%%%%%%%%%%%%%%%%%%%%%%%%
\subsubsection{The case of nonlinear potential bias and temperature gradient}
%%%%%%%%%%%%%%%%%%%%%%%%%%%%%%%%%%%%%%%%%%%%%%%%%%%%%%

The nonequilibrium thermoelectric properties of the quantum dot coupled
to both ferromagnetic leads are shown in \fig{fig:nonlin_P}.
The first row corresponds to the case of parallel configuration of the leads' magnetizations.
The differential conductance depicted in \fig{fig:nonlin_P}(a)
exhibits the split Kondo anomaly, with side peaks of similar magnitude
located at roughly the same distance from the zero bias.
Both peaks die off with the temperature gradient around $\Delta T \approx 0.05 \, \G_L$,
i.e. when thermal gradient exceeds the Kondo temperature.

At low $\Delta T$ the differential and nonequilibrium
Seebeck coefficients exhibit similar bias voltage dependence 
to the case presented in Figs.~\ref{fig:nonlin_NM}(e) and (f),
see Figs.~\ref{fig:nonlin_P}(b) and (c). Now, however, the region
of negative Seebeck coefficient is smaller. This can be attributed to
the fact that the split Kondo resonance is more symmetric
across the bias reversal in the case of parallel magnetic configuration,
cf. Fig.~\ref{fig:G}(b). Unlike in the case of nonmagnetic right lead,
the sign changes at finite bias corresponding to the split Kondo peak
persist as long as $\Delta T \lesssim T_K$ and 
disappear around comparable temperature gradient.

The case of antiparallel magnetic configuration of the system
is presented in the second row of \fig{fig:nonlin_P}.
Consistent with the discussion of \fig{fig:G}(c),
the differential conductance exhibits two conductance peaks
but with a large difference in their magnitudes.
The peak in the negative bias regime is far more pronounced
than the miniscule peak one can observe in the positive regime.
Just as in the case of other configurations,
the peaks die out with increasing the temperature
gradient but the negative bias peak survives
till larger temperature gradients $\Delta T \approx 0.2 \G_L$
whereas the positive bias peak vanishes
at temperature gradients as low as $\Delta T \approx 0.02 \G_L$.

The Seebeck coefficients $S_d^{AP}$ and $S_n^{AP}$,
shown in Figs.~\ref{fig:nonlin_P}(e) and (f), respectively,
demonstrate a similar behavior to the other configurations
only at very low temperature gradients.
However, now, instead of sign changes,
one only observes suppression
of the Seebeck coefficients at the
corresponding values of the bias voltage
associated with the exchange field.
These suppressions extend to temperatures gradients
of the order of $\Delta T \approx 0.03 \G_L$,
see Figs.~\ref{fig:nonlin_P}(e) and (f).

%%%%%%%%%%%%%%%%%%%%%%%%%%%%%%%%%%%%%%%%%%%%%%%%%%%%%%%%%%%%%

\begin{figure*}[t]
	\includegraphics[width=0.9\textwidth]{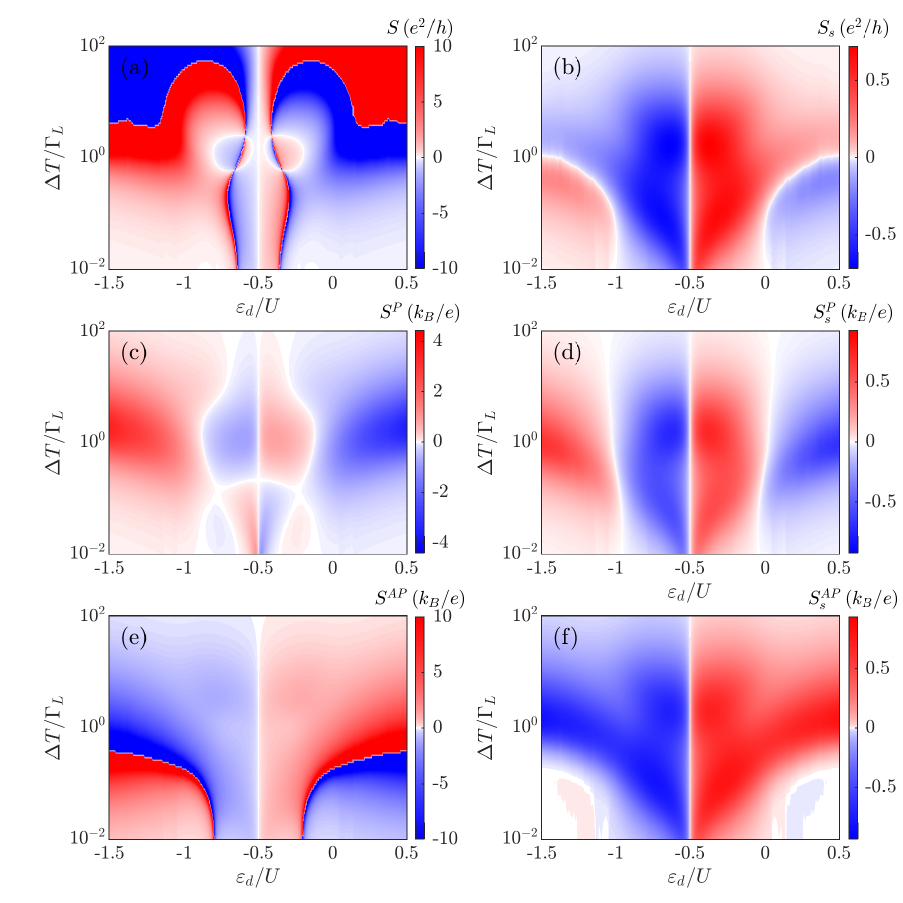}
	\caption{The charge Seebeck (first column) and the spin Seebeck (second column)
		coefficients under nonlinear temperature gradient $\Delta T$
		and linear response spin bias $V_s$ as a function of the orbital level energy $\e_d$ and $\Delta T$.
		The first row corresponds
		to the case of nonmagnetic right lead, while
		the second (third) row presents the case of ferromagnetic right lead
		in the parallel (antiparallel) magnetic configuration of the system.
		The other parameters are the same as in \fig{fig:lin_P}.
	}
	\label{fig:lin_Ss}
\end{figure*}

%%%%%%%%%%%%%%%%%%%%%%%%%%%%%%%%%%%%%%%%%%%%%%%%%%%%%%
\subsection{Finite spin accumulation and the associated nonequilibrium spin Seebeck effect}
%%%%%%%%%%%%%%%%%%%%%%%%%%%%%%%%%%%%%%%%%%%%%%%%%%%%%%

In this section we consider the case when ferromagnetic contacts 
are characterized by slow spin relaxation, which can result in a finite spin accumulation \cite{Swirkowicz2009Aug,Swirkowicz2009Nov}.
Such a spin accumulation will induce a spin bias across the quantum dot.
Here, we assume that the spin accumulation and the resulting spin-dependent
chemical potential occurs only in the right lead.
Thus, we define the induced spin bias as, $V_{s}/2=\mu_{R\up}=-\mu_{R\down}$ (keeping $\mu_L=0$).
The nonequilibrium spin bias across the quantum dot enables the spin chemical potentials
to be tuned separately and thus the thermal bias induced transport can be different in the separate spin channels.
The system can then exhibit interesting spin caloritronic properties,
such as the spin Seebeck effect in this setup.
The spin Seebeck coefficient $S_s$ quantifies the magnitude and the direction of
the spin current induced in the presence of a thermal bias \cite{Uchida2008Oct}.
Analogous to the differential Seebeck effect $S_d$,
the differential spin Seebeck coefficient $S_s$ in the nonlinear response regime can be defined as
\be
S_{s}= -\left(\frac{dV_{S}}{d \Delta T}\right)_{\!\! I_s}
= -\left(\frac{\partial I_{s}}{\partial \Delta T}\right)_{\!\! V_s} \bigg/\! \left(\frac{\partial I_{s}}{\partial V_s}\right)_{\!\! \Delta T} ,
\label{Eq:S_s}
\ee
where $I_s = I_{\up}(\mu_{R\up},\Delta T)-I_{\down}(\mu_{R\down},\Delta T)$ is the net spin current flowing through the system.
This quantity acts as a response over the spin current as a function of
both the spin bias $V_s$ and the temperature gradient $\Delta T$.
In addition to the net spin current, there can also exist a charge current
$I = \sum_{\s} I_{\s}(\mu_{R\s},\Delta T)$ flowing across the system
originating solely from the thermal and the spin biases.
We define the Seebeck coefficient that estimates the charge current in the presence
of the spin bias as the charge Seebeck coefficient $S$
\cite{Swirkowicz2009Aug}.
The charge Seebeck coefficient $S$ can thus be defined
based on the response of charge current $I$ as
\be
S= -\left(\frac{dV_{s}}{d \Delta T}\right)_{\!\! I} 
= -\left(\frac{\partial I}{\partial \Delta T}\right)_{\!\! V_s} \bigg/\! \left(\frac{\partial I}{\partial V_s}\right)_{\!\! \Delta T}.
\label{Eq:S_res}
\ee

We first discuss the case of linear response in the spin bias $V_s$
with large and finite temperature gradient $\Delta T$,
focusing on the differential spin Seebeck coefficient $S_s$
and the charge Seebeck coefficient $S$.
It is pertinent to note that the nonequilibrium equivalent of the spin Seebeck coefficient $S_{s,n}$
tends to remain undefined in our considerations,
since the magnitude of the spin bias fails to compensate
for the thermally induced spin current in (\textit{parts of}) the regimes considered.
Hence in this paper, we limit our discussions to the differential spin Seebeck coefficient
$S_{s}\equiv S_{s,d}$ in the case of different configurations.
We further investigate the dependence of $S_s$ and $S$
on large and finite spin bias under applied temperature gradient.

\subsubsection{The case of zero spin bias with nonlinear temperature gradient}

Figure \ref{fig:lin_Ss} shows the behavior of the charge Seebeck coefficients $S$, $S^P$, $S^{AP}$
and the spin Seebeck coefficients $S_s$, $S_s^{P}$, $S_s^{AP}$
for the case of nonmagnetic right lead, as well as the case of ferromagnetic lead in the parallel and antiparallel magnetic configurations, respectively.
The first row of \Fig{fig:lin_Ss} shows the case of right lead with spin polarization $p=0$,
but with finite spin accumulation occurring from the spin-resolved transport through the quantum dot.
Figure \ref{fig:lin_Ss}(a) displays the charge Seebeck $S$ coefficient,
which behaves similarly to the differential Seebeck $S_d$ presented in \Fig{fig:lin_NM}
except some points of divergences.
At temperature gradients smaller than $\Gamma_L$,
there exist two additional sign changes,
both in the local moment regime symmetric across the particle-hole symmetry point.
The points of sign change spread out of the local moment regime
for thermal biases $\Delta T\gtrsim  3\, \Gamma_L$.
The sign changes of the Seebeck effect are also accompanied by large divergences in the magnitude of $S$.
The additional sign changes and divergences originate from the behavior of the denominator
in the definition of $S$, cf. \Eq{Eq:S_res}.
The denominator in \Eq{Eq:S_res}, which can be represented as,
$G^{cs}=(\partial I/\partial V_s)_{\Delta T}$,
is the differential mixed conductance \cite{Swirkowicz2009Aug}
that estimates the charge current in the presence of a spin bias,
which can be either negative or positive, resulting in its zero crossing points causing the divergence.
From a physical perspective, tuning the temperature gradient in these specific regimes
will result in extraordinary changes in the induced charge current.
Note that the colormaps in Figs.~\ref{fig:lin_Ss}(a) and (e) have been truncated for readability.

The charge Seebeck for the parallel configuration
[see \Fig{fig:lin_Ss}(c)] perfectly recreates the behavior seen in \Fig{fig:lin_P}(c).
In the case of the parallel configuration, the relative scaling of the couplings
in each spin channels on the right and left is the same,
resulting in a non-negative $G^{cs}$ and, thus, no divergences.
Similarly to \Fig{fig:lin_Ss}(e), the charge Seebeck effect for the antiparallel configuration,
there exist a resemblance to the Seebeck coefficient discussed in \Fig{fig:lin_P}(d),
but overlaid by the divergences associated with $G^{cs}$.
In this case, the additional sign changes start from inside the local moment regime
at very low temperature gradients and move out of the local moment regime
monotonously around $\Delta T \approx 10^{-1}\, \Gamma_L$. 

The differential spin Seebeck coefficient $S_s$
shown in panels (b), (d) and (f) of Fig.~\ref{fig:lin_Ss} for different lead configurations
behave antisymmetrically across the particle-hole symmetry point ($\e_d=-U/2$).
There exist a pronounced spin Seebeck coefficient in the local moment regime
for all the configurations that dies off at $\Delta T\gtrsim 10\, \Gamma_L$.
Such regions of considerable spin Seebeck effect
have been observed in the linear response studies of symmetrically
coupled quantum dots as a function of the global temperature $T$ \cite{Weymann2013Aug,Manaparambil2021Apr}.
In addition to the sign change at the particle-hole symmetry point,
at very low $\Delta T$ $S_s$ changes sign when moving out of the local moment regime (i.e., at $\e_d\approx -U,0$).
In the case of the nonmagnetic right lead,
the region of sign change outside the local moment regime extends up to $\Delta T \approx \G_L$,
whereas for the antiparallel configuration the sign change extends only up to $\Delta T\approx 0.2\,\G_L$.
On the other hand, the sign change of the spin Seebeck in the local moment regime
survives at thermal gradients even greater than $\Delta T \approx 10^2\,\G_L$ for the parallel configuration.

\subsubsection{The case of nonlinear spin bias and temperature gradient}

\begin{figure*}[t]
	\includegraphics[width=0.9\textwidth]{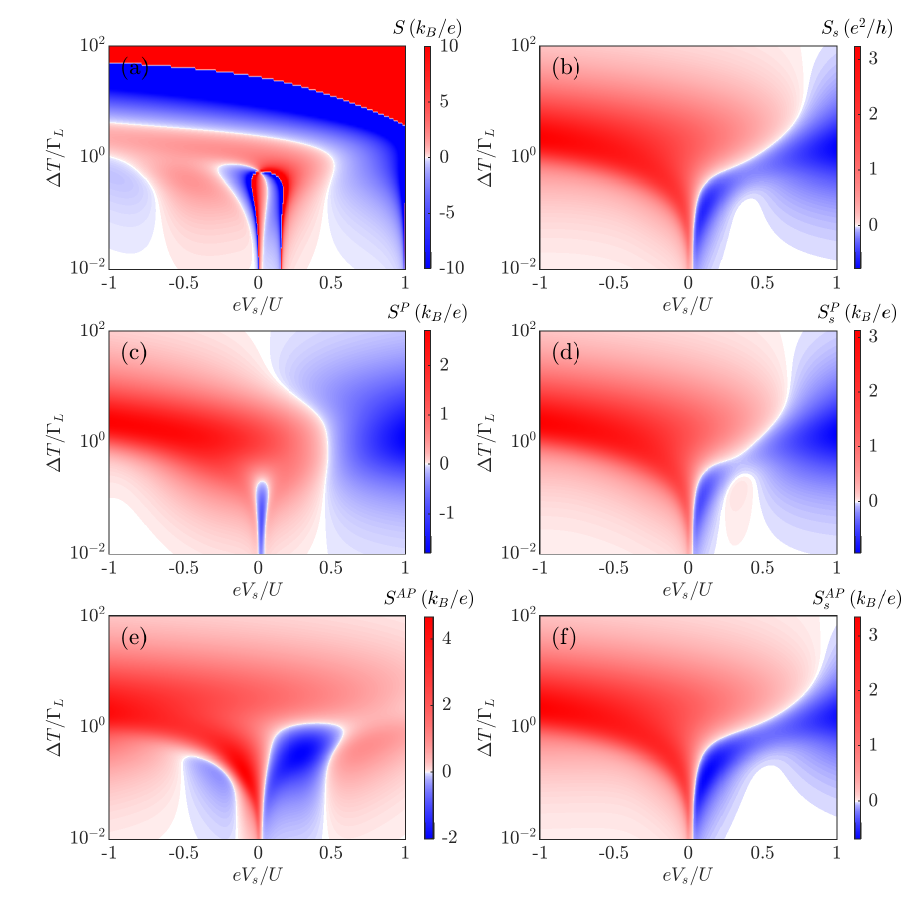}
	\caption{
	The charge Seebeck (first column) and the spin Seebeck (second column)
	coefficients for the orbital level $\e_d=-U/3$ as a function of the
	applied spin bias $V_s$ and $\Delta T$.
	The first row corresponds
	to the case of nonmagnetic right lead, while
	the second (third) row presents the case of ferromagnetic right lead
	in the parallel (antiparallel) magnetic configuration of the system.
	The other parameters are the same as in \fig{fig:lin_P}.}
	\label{fig:nonlin_Ss}
\end{figure*}

The dependence of the nonlinear charge Seebeck
and the spin Seebeck effect is shown in \Fig{fig:nonlin_Ss} for the case of orbital energy level $\e_d=-U/3$.
The first column in \Fig{fig:nonlin_Ss} focuses on the charge Seebeck effect
for various magnetic configurations of the system.
For the case of a nonmagnetic right lead,
the charge Seebeck coefficient $S$ changes sign four times as a function of $V_s$
at temperature gradients below $\Delta T\approx 0.5\G_L$, see \Fig{fig:nonlin_Ss}(a).
Two among these sign changes (around $V_s\approx 0.001\, U$ and $V_s \approx 0.15\, U$)
correspond to the zeros in the mixed conductance $G^{cs}$,
which can be identified from the divergence in $S$ around the sign changes.
The other two sign changes (around $V_s\approx-0.05\, U$ and $V_s\approx 0.03\, U$)
originate from the zeros of the thermal response
$-({\partial I}/{\partial \Delta T})_{V_s}$,
i.e. the numerator in the definition of the charge Seebeck, cf. \Eq{Eq:S_s}.
As the temperature gradient increases, the regions of sign change introduced by $G^cs$
and the temperature gradient become larger in the spin bias regime
until around $\Delta T\approx \G_L/3$ for the sign change
associated with the mixed conductance and $\Delta T\approx \G_L/2$
due to the sign change from the thermal response.
With further increase in the temperature gradient $\Delta T$
the regions of sign change disappear. This happens around $\Delta T\gtrsim \G_L/2$
for the sign change caused by the mixed conductance
and $\Delta T\gtrsim 0.8\G_L$ for the sign change due to the thermal response.
The remaining two sign changes at $V_s=-U/2$ and $V_s=U/2$ correspond to the Hubbard peaks of the quantum dot spectral function.
The region of these sign changes disappears above temperature gradient $\Delta T \gtrsim 4\,\G_L$.
At $V_s=0$ and very large temperature gradients (around $\Delta T\gtrsim 10\, \G_L$),
there exist another sign change that originates from the zeros of $G^{cs}$.
For positive $V_s$, this sign change moves to lower $\Delta T$,
while for negative $V_s$ this sign change moves to higher $\Delta T$, see \Fig{fig:nonlin_Ss}(a).

Figure~\ref{fig:nonlin_Ss}(c) shows the charge Seebeck effect $S^P$
corresponding to the system in parallel configuration of the leads.
We observe that there are two sign changes as a function of the spin bias $V_s$.
At low temperatures, $\Delta T\lesssim 0.01\, \G_L$,
the region of sign change appears between $V_s \approx 0.005\, U$ and $V_s \approx U/2$.
One can identify that these sign changes originate solely from the thermal response of the current under spin bias.
With an increase in $\Delta T$, the sign change at $V_s \approx 0.005\, U$
crosses over to the negative $V_s$ regime
and the sign change around $V\approx U/2$ moves closer to $V_s\approx 2\,U/3$,
thus increasing the region of sign change in the spin bias $V_s$ regime
until around $\Delta T \approx 0.1\, G_L$.
On further increase in temperature gradient, the regions of sign change
tend to disappear once $\Delta T \approx 0.2\,G_L$.
On the other hand, outside of this regime, the sign of the spin-resolved thermopower remains positive.
We also note that there exist another point of sign change due to the contribution from
the Hubbard peaks in the spectral function.
Unlike in the previous case of $S$, this sign change survives for
large temperature gradients $\Delta T$ and moves closer to $V_s=0$
when the temperature gradient $\Delta T$ is increased $\Delta T \gtrsim \G_L$, see Fig.~\ref{fig:nonlin_Ss}(c).

The charge Seebeck coefficient for the antiparallel configuration $S^{AP}$
does not show any sign change in the local moment regime apart from the particle-hole symmetry point $\e_d=-U/2$,
as seen in \Fig{fig:lin_Ss}(e). However, as a function of the spin bias $V_s$,
two points of sign changes form in the dependence of the charge Seebeck effect $S^{AP}$.
One change occurs in the negative spin bias regime around $V_s\approx -0.15\, U$
and the other one in the positive regime at $V_s\approx 0.03\, U$.
With increasing $\Delta T$, these changes move further apart into the negative and positive spin bias regimes, respectively.

It is important to emphasize that the sign changes observed
in the charge Seebeck coefficient as a function of spin bias $V_s$
do not correspond to the sign changes seen in the Seebeck coefficient
as a function of $V$, as discussed and presented in \Fig{fig:nonlin_NM} and \Fig{fig:nonlin_P}.
This is associated with the fact that the generated current [\Eq{Eq:I}]
as a function of $V$  scans through each of the split Kondo resonances
shown in \Fig{FigA} separately, resulting in the split peaks
seen in the differential conductance and the corresponding sign changes in the Seebeck coefficients.
However, as a function of the spin bias $V_s$, the signatures from the split Kondo resonance
cannot be identified directly in the generated current $I$.
This is because the spin bias $\mu_{R\up}-\mu_{R\down}=V_s$ scans both split Kondo peaks
(see \Fig{FigA}) simultaneously, and the total current $I$ is rescaled
by just relative couplings of the separate spin channels $\G_{R\s}$.
Hence, the sign changes in the charge Seebeck coefficient
are solely resulting from the sign changes in the thermal response and the mixed charge conductance.

The spin Seebeck coefficient in the nonlinear spin bias regime
is presented in the second column of \Fig{fig:nonlin_Ss}.
Panels (b),(d) and (f) show the case of the nonmagnetic right lead as well as ferromagnetic right lead in
the parallel and antiparallel configuration, respectively.
From the discussion of the linear $V_s$ case shown in \Fig{fig:lin_Ss},
we observe that the differential spin Seebeck coefficient does not change
sign inside the local moment regime for all three configurations.
Under finite spin bias $V_s$, we can see only one sign change in the positive
spin bias regime around $V\approx U/4$ for all magnetic configurations.
The point of sign change shifts towards the positive regime with increasing temperature gradient $\Delta T$.
The behavior of the spin Seebeck coefficient is identical for all the configurations
apart from slight differences in the magnitude,
meaning that this originates solely from the properties of the spectral function outside the split Kondo peaks. 
In the case of parallel configuration, we observe a small region of additional
sign change around $V_s\approx 0.2\, U$ to $V_s\approx U/2$.
Such a behavior have already been observed in the nonequilibrium thermopower
of similar systems where it has been attributed to the characteristic behavior of the spectral function
for energies between the Kondo and the Hubbard peak \cite{Manaparambil2023Feb}.

%%%%%%%%%%%%%%%%%%%%%%%%%%%%%%%%%%%%%%%%%%%%%%%%%%
\section{\label{sec:summary}Summary}
%%%%%%%%%%%%%%%%%%%%%%%%%%%%%%%%%%%%%%%%%%%%%%%%%%

In this paper we have studied the nonequilibrium thermoelectric properties
of the system consisting of a quantum dot/molecule asymmetrically coupled to
external ferromagnetic leads.
The strongly coupled ferromagnetic contact induces
an exchange field in the dot that can split and suppress the Kondo resonance.
The emphasis has been put on the signatures
of the interplay between spin-resolved tunneling
and strong electron correlations in the nonequilibrium 
thermopower of the system. In particular, we have determined
the bias voltage and temperature gradient dependence of 
the differential and nonequilibrium Seebeck coefficients.
We have observed new signatures in the Seebeck coefficients corresponding to the
Kondo resonance and the regions where the ferromagnetic contact induced
exchange field suppresses the Kondo effect both
in the potential bias and temperature gradient regimes.
More specifically, we have demonstrated that 
the Seebeck coefficient exhibits new sign changes
as a function of bias voltage, which are associated
with the split Kondo resonance. These sign changes
extend to the temperature gradients on the order of the Kondo temperature.
Furthermore, we investigated the influence of the spin accumulation and the resulting spin bias
on the Seebeck and spin Seebeck coefficients.
The nonlinear charge Seebeck coefficient and the spin Seebeck coefficient
showed points of sign changes in the presence of finite spin and thermal bias,
corresponding to the different properties of the quantum dot spectral function.

%%%%%%%%%%%%%%%%%%%%%%%%%%%%%%%%%%%%%%%%%%%%%%%%%%%%%%%%%%%%%%%%
%%%%%%%%%%%%%%%%%%%%%%%%%%%%%%%%%%%%%%%%%%%%%%%%%%%%%%%%%%%%%%%%

\begin{acknowledgments}
	This work was supported by the Polish National Science
	Centre from funds awarded through the decision No. 2017/27/B/ST3/00621.
	We also acknowledge the computing time
	at the Pozna\'{n} Supercomputing and Networking Center. 
\end{acknowledgments}

%%%%%%%%%%%%%%%%%%%%%%%%%%%%%%%%%%%%%%%%%%%%%%%%%%%%%%
%%%%%%%%%%%%%%%%%%%%%%%%%%%%%%%%%%%%%%%%%%%%%%%%%%%%%%
%apsrev4-2.bst 2019-01-14 (MD) hand-edited version of apsrev4-1.bst
%Control: key (0)
%Control: author (8) initials jnrlst
%Control: editor formatted (1) identically to author
%Control: production of article title (0) allowed
%Control: page (0) single
%Control: year (1) truncated
%Control: production of eprint (0) enabled
%

%%%%%%%%%%%%%%%%%%%%%%%%%%%%%%%%%%%%%%%%%%%%%%%%%%%%%%
%%%%%%%%%%%%%%%%%%%%%%%%%%%%%%%%%%%%%%%%%%%%%%%%%%%%%%

\end{document}